\documentclass[a4paper,fleqn,usenatbib]{mnras}

\usepackage{newtxtext,newtxmath}

\usepackage[T1]{fontenc}
\usepackage{ae,aecompl}

\usepackage{graphicx}
\usepackage{amsmath}
\usepackage{amssymb}

\title[Three-year WISE/NEOWISE Coadds]{Deep Full-sky Coadds from Three Years of WISE and NEOWISE Observations}

\author[Meisner et al.]{
A.~M. Meisner,$^{1,2}$\thanks{ameisner@lbl.gov}
D. Lang$^{3}$
and D.~J. Schlegel$^{2}$
\\
$^{1}$Berkeley Center for Cosmological Physics, Berkeley, CA 94720, USA \\
$^{2}$Lawrence Berkeley National Laboratory, Berkeley, CA, 94720, USA \\
$^{3}$Department of Astronomy \& Astrophysics and Dunlap Institute, University of Toronto, Toronto, ON M5S 3H4, Canada \\
$^{4}$Department of Physics \& Astronomy, University of Waterloo, 200 University Avenue West, Waterloo, ON, N2L 3G1, Canada
}

\pubyear{2016}

\begin{document}
\label{firstpage}
\pagerange{\pageref{firstpage}--\pageref{lastpage}}
\maketitle

\begin{abstract}
We have reprocessed over 100 terabytes of single-exposure WISE/NEOWISE images to create the deepest ever full-sky maps at 3$-$5 microns. We incoporate all publicly available W1 and W2 imaging -- a total of $\sim$8 million exposures in each band -- from $\sim$37 months of observations spanning 2010 January to 2015 December. Our coadds preserve the native WISE resolution and feature depth of coverage $\sim$3$\times$ greater than that of the AllWISE Atlas stacks. Our coadds are designed to enable deep forced photometry, in particular for the  Dark Energy Camera Legacy Survey (DECaLS) and Mayall z-Band Legacy Survey (MzLS), both of which are being used to select targets for the Dark Energy Spectroscopic Instrument (DESI). We describe newly introduced processing steps aimed at leveraging added redundancy to remove artifacts, with the intent of facilitating uniform target selection and searches for rare/exotic objects (e.g. high-redshift quasars and distant galaxy clusters). Forced photometry depths achieved with these coadds extend 0.56 (0.46) magnitudes deeper in W1 (W2) than is possible with only pre-hibernation WISE imaging.
\end{abstract}


\section{Introduction}

The Wide-field Infrared Survey Explorer \citep[WISE; ][]{wright10} was designed to map the entire
sky at mid-infrared wavelengths with sensitivity far exceeding that of its predecessors, IRAS \citep{iras} and DIRBE \citep{dirbe}. As a source of high-quality, full-sky imaging, its observations have an extremely wide range of applications, from near-Earth asteroids \citep[e.g.][]{2010TK7} to the most luminous galaxies in the  universe \citep[e.g.][]{tsai15}.

WISE is a 0.4 meter telescope onboard a satellite in low-Earth orbit, launched in late 2009. Always pointing near 90$^{\circ}$ solar elongation while making successive scans in ecliptic latitude at fixed ecliptic longitude, WISE maps the entire sky once every six months when operational. From 2010 January to 2010 August, WISE carried out a full-sky mapping in each of four broad mid-infrared
bandpasses centered at 3.4$\mu$m (W1), 4.6$\mu$m (W2), 12$\mu$m (W3) and 22$\mu$m (W4).

Due to the depletion of solid hydrogen cryogen, the W3 and W4 channels were
rendered unusable as of 2010 September and 2010 August, respectively. However, WISE continued surveying in W1 and W2
through  2011 January as part of the Near-Earth Object Wide-field Infrared Survey Explorer \citep[NEOWISE; ][]{neowise} mission. In 2011 February WISE was placed in hibernation, temporarily ceasing data acquisition. In December 2013, WISE recommenced surveying the sky in W1 and W2, carrying out the NEOWISE-Reactivation \citep[NEOWISER; ][]{neowiser} mission. Despite the multi-year hiatus, NEOWISER images are of essentially the same high quality and sensitivity as exposures acquired pre-hibernation \citep{neowiser}. In March 2015, NEOWISER published its first year's worth of single-exposure images and frame-level source extractions. In March 2016, a second year of such data were made
public.

Because NEOWISER is an asteroid hunting and characterization mission, it does not deliver 
any coadded data products. Nevertheless, the astrophysics research community has recognized
the tremendous value of coadded data products incorporating NEOWISER images \cite[e.g.][]{faherty}.

The Dark Energy Spectroscopic Instrument \citep[DESI, ][]{desi, desi_part1, desi_part2} represents an important application which stands to benefit from access to deep WISE stacks, and which ultimately drives many of our design considerations in building coadded WISE/NEOWISE data
products. DESI will select millions of luminous red galaxy (LRG) targets from $r$+$z$+W1
photometry, while its quasar targeting will additionally make use of $g$ and W2 fluxes.  To obtain high-quality photometrically selected samples of these targets, DESI requires the deepest and cleanest possible W1/W2 coadds.

DESI targeting employs a `forced photometry' approach which measures WISE fluxes for \textit{every}
optically detected source, fixing the centroid and morphology to those obtained from much
higher resolution optical imaging. This technique has already been successfully applied to 
optical catalogs from SDSS \citep{lang14b} and DECam \citep{decals, meisner16}. The
\cite{lang14b} results have played a significant role in eBOSS LRG and quasar selection \citep{eboss_lrg, prakash15, eboss_qso}.

In the eBOSS/DESI targeting applications, WISE forced photometry has been performed on `unWISE' coadds \citep{lang14, meisner16}. The original unWISE coadds of \cite{lang14} used the available WISE data from observations conducted in 2010 and early 2011. The W1/W2 coadds of \cite{meisner16} are based on an adaptation of the \cite{lang14} unWISE coaddition pipeline, folding in the first year of NEOWISER imaging. \cite{meisner16} thereby doubled the depth of coverage while eliminating the dominant artifacts found in the \cite{lang14} coadds.

Here we update the results of \cite{meisner16}, adding in the most recently published year of NEOWISER W1/W2 exposures. We highlight our recent processing improvements which
aid in the elimination and flagging of remaining artifacts. Our latest coadds represent
a 50\% increase in depth of coverage relative to those of \cite{meisner16}, and a 200\% increase
relative to those of \cite{lang14} in W1 and W2.

In $\S$\ref{sec:data} we list the input data used for this work. In $\S$\ref{sec:unwise} we provide an 
overview of unWISE coaddition and our image processing strategy/philosophy. In $\S$\ref{sec:updates} we highlight newly added improvements to our coaddition pipeline. In $\S$\ref{sec:results} we display and validate key aspects of our new W1/W2 full-sky maps. We conclude in $\S$\ref{sec:conclusion}.

\begin{figure*} 
	\includegraphics[width=7.0in]{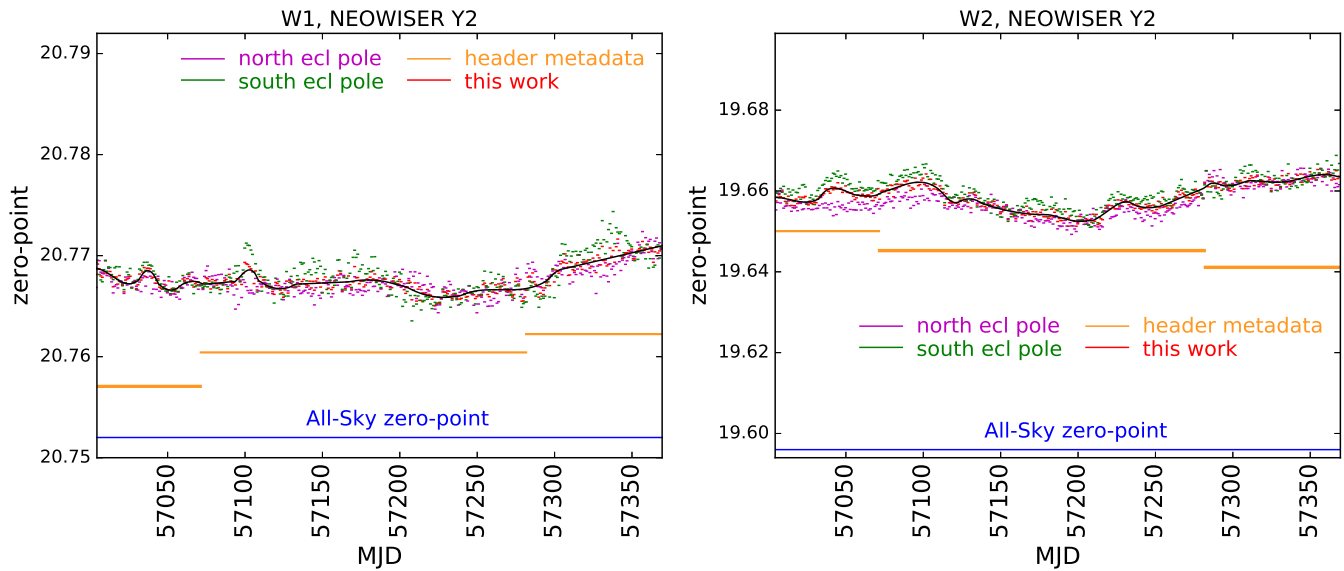}
    \caption{Custom photometric zero-points derived for the second year of NEOWISER data according to the procedure described in $\S$4 of \citet{meisner16}. Left: W1. Right: W2. Our
per-day zero-point measurements are shown as red dashes. Black lines show the
smooth functions used to interpolate off of the per-day zero-point measurements during coaddition. There is no discontinuity in our derived zero-points at the boundary between the first and second year
NEOWISER releases (MJD = 57004.3, see Figure 1 of \citet{meisner16} for comparison).}
    \label{fig:zp}
\end{figure*}

\section{Data}
\label{sec:data}
Our coaddition proceeds from the least-processed form of publicly available WISE imaging, namely the
``Level 1b" (L1b) single-exposure framesets. We downloaded a local copy of every publicly available W1 and W2 frameset, including those from the All-Sky, 3band Cryo, NEOWISE, NEOWISER year one (NEO1) and NEOWISER year two (NEO2) releases \citep{cutri12, cutri13, cutri15}. For each frameset, a \verb|-int-| FITS file gives the sky intensity, while a \verb|-unc-| FITS file provides per-pixel uncertainty estimates, and a \verb|-msk-| FITS file contains a bitmask flagging artifacts such as bad pixels and cosmic rays. In all, we downloaded $\sim$52 TB of L1b data products per band, totaling 49 terapixels of inputs.

To validate photometry derived from our coadds, we make use of the AllWISE Source Catalog \citep{cutri13}, as
well as W1/W2 forced photometry from \cite{lang14b} and Data Release 4 (DR4) of the DESI imaging Legacy Survey\footnote{\url{http://legacysurvey.org}. DR4 is currently internal to the DESI collaboration, but
is expected to become public in the near future.}.

\section{Processing Overview}
\label{sec:unwise}

\subsection{Atlas versus unWISE Coadds}
There are two main types of full-sky coadded WISE data products presently available. First, there
are the Atlas stacks created by the WISE/NEOWISE team. These images are intentionally blurred
by the WISE point spread function (PSF), since their primary purpose is to enable deep source extraction for e.g. the AllWISE Source Catalog\footnote{Once detections are in hand, however, the AllWISE catalog measurements for each source are derived via joint modeling of all relevant single-exposure images.}. The most recent full-sky set of Atlas coadds is that of the AllWISE release and is based solely on pre-hibernation exposures.

Alternatively, the unWISE \citep{lang14} line of coadds uses Lanczos interpolation to preserve the native WISE resolution. The unWISE stacks are designed for forced photometry, a use case for 
which any blurring of the native WISE PSF would be suboptimal. This work is part of an ongoing effort
to upgrade the W1/W2 unWISE coadds with each new release of NEOWISER L1b
exposures, enhancing the depths achieved and using added redundancy to remove artifacts.

\subsection{unWISE Philosophy}

unWISE coaddition is meant to be lightweight/minimalist, with the plurality of compute time 
consumed by reading in the large number of L1b frames which overlap a given coadd footprint, and
subsequently writing the output coadd images. As such, no truly robust outlier rejection methods 
(e.g. median filtering the resampled L1b intensities at each coadd pixel location) are employed, as
these would dramatically increase the total computational cost.

Given that we have already produced full-depth coadds including all exposures through the
first-year NEOWISER release \citep{meisner16}, one might imagine a scheme in which we
update our coadds by \textit{only} processing the ``new'' exposures, adding them onto the existing coadd. Such an approach would not fully take advantage of improved outlier rejection enabled by the two newly added sky passes from second-year NEOWISER observations. For instance, we would inherit artifacts present in previous iterations of the unWISE coadds, rather than leveraging the newly added redundancy of the latest set of exposures to improve artifact removal (see e.g. Figure \ref{fig:moon}). Such an approach would also likely require us to implement a series of contrived shortcuts/hacks. 

Therefore, \textit{with every new release of NEOWISER exposures, we opt to rebuild all of our
full-depth coadds from scratch}. Even when jointly processing all exposures spanning
2010 January to 2015 December, as we have chosen to do, the total full-sky computational 
expense is only of order tens of thousands of CPU hours.

In combining observations which span the full WISE lifetime, we also strive to achieve the best possible
relative calibration of all exposures. To accomplish this, we adopt a custom photometric
calibration using repeat measurements of calibrator sources near the ecliptic poles, as described in 
$\S$4 of \cite{meisner16}. We have now performed this photometric calibration analysis for the second year of NEOWISER observations. The results are shown in Figure \ref{fig:zp}.

One final component of our philosophy in generating WISE coadds is to throw away as 
little data as possible. This motivated our work in \cite{meisner16} to recover 
exposures corrupted by scattered moonlight. In this same vein, we retain all frames at/near the ecliptic
poles, despite the added computational cost associated with exceptionally large depth of coverage
in these regions. As a result, the tiles at the north/south ecliptic poles have peak integer coverage of $\sim$17,000 frames per band, for a total integration time of $\sim$1.5 days.

\subsection{unWISE Implementation Details}

unWISE coadds employ the same set of 18,240 tile centers as do the  AllWISE Atlas stacks. We adopt
a pixel scale matching that of the W1/W2 L1b exposures themselves, 2.75$''$ per pixel. Each coadd
astrometric footprint has its $-x$ ($+y$) axis aligned with celestial east (north). We refer to
a coadd astrometric footprint as a `tile', and identify tiles by their \verb|coadd_id| values, which are
strings encoding their central (RA, Dec) coordinates, e.g. `0000p000'. Our coadds are 1.56$^{\circ}$ (2048 pixels) on a side.

unWISE coaddition proceeds by identifying all exposures that overlap the tile of interest, 
then resampling every relevant exposure onto the coadd pixel grid using Lanczos interpolation.
With all resampled single-exposure pixels in memory, a ``first round'' of coaddition computes
a simple mean and standard deviation of the contributing L1b intensity values at each coadd pixel location. These first-round coadd mean and standard deviation images are then used 
to enable outlier rejection during a second and final round of coaddition. For full details refer to \cite{lang14} and \cite{meisner16}.

\section{Coaddition Updates/Improvements}
\label{sec:updates}

The foremost improvement achieved in this work relative to the coadds of \cite{meisner16} 
is the $\sim$50\% enhancement in depth of coverage obtained by including second-year NEOWISER
observations. However, because
we uniformly reprocess all exposures following every NEOWISER release, each new, deeper set of coadds also represents an opportunity
to build further processing improvements into the unWISE coaddition pipeline. The following
subsections describe newly introduced facets of the unWISE processing and data products.

\subsection{Min/Max Rejection}
\begin{figure*}
	\includegraphics[width=7.0in]{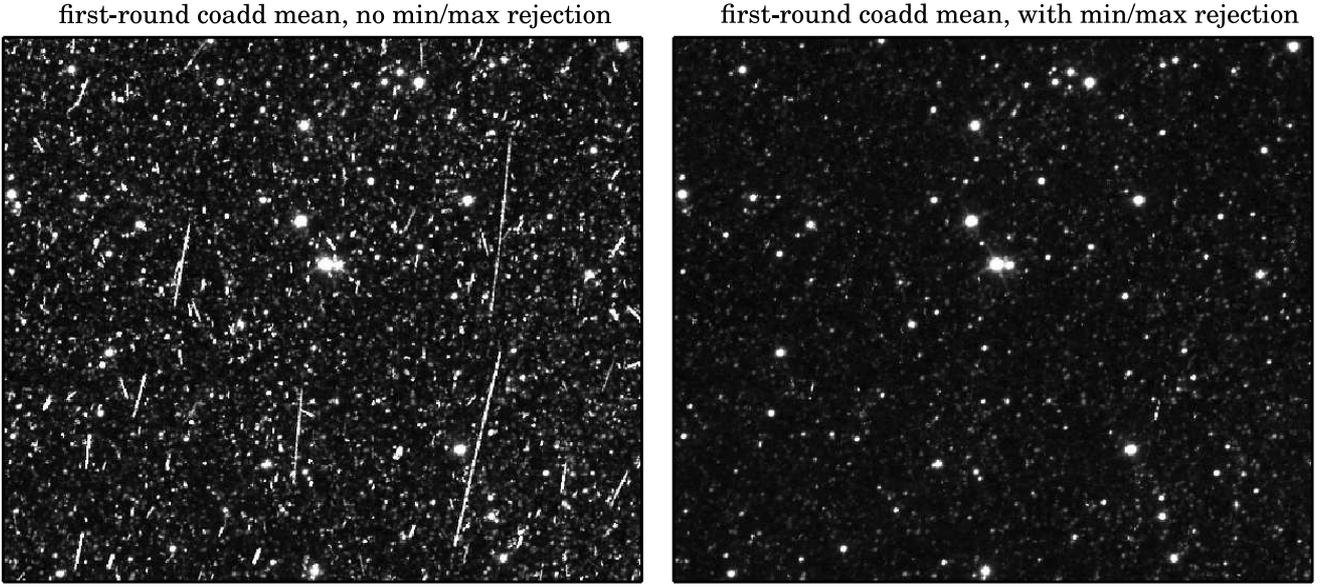}
    \caption{Benefits of min/max rejection for a $22.9' \times 19.7' $  first-round coadd region affected by the 
    South Atlantic Anomaly, drawn from tile 0478m243 and centered at ($\alpha$, $\delta$) = 
    (48.536$^{\circ}$, $-23.843^{\circ}$). Left: W1 first-round coadd
    mean image accumulated without min/max rejection step. Right: W1 first-round coadd
    mean image accumulated with newly implemented min/max rejection step. When the first-round
    coadd mean is accumulated by simple averaging, as in \citet{lang14} and \citet{meisner16}, it
    is severely contaminated by cosmic rays. Removing the minimum and maximum single-exposure
    pixel values at each first-round coadd pixel location dramatically reduces the imprint of cosmic rays,
    leading to improved outlier rejection during the second round of coaddition.}
    \label{fig:min_max_r1}
\end{figure*}

As mentioned previously, unWISE coaddition does not use robust statistics to perform outlier
rejection, instead relying on naive first-round coadd mean and standard deviation images to flag
problematic pixels in the L1b exposures. This computationally cheap approach is vulnerable to the appearance of
extreme outliers in a small number of exposures. The most conspicuous example of this 
phenomenon arises due to large numbers of cosmic ray strikes in frames affected by the South 
Atlantic Anomaly, which mainly impacts observations at $-45^{\circ} < \delta < -10^{\circ}$. This leads to first-round coadd results which are strongly contaminated
by cosmic rays, and therefore are relatively ineffective at flagging outliers (see 
left panel of Figure \ref{fig:min_max_r1}).

To combat such corruption of our first-round coadds, we have now implemented a ``min/max''
rejection step during first-round coaddition. Specifically, for each pixel in coadd space, we 
precompute which L1b exposures contribute the highest and lowest single-frame intensity values at that
location. Then, when computing the first round coadd mean and standard deviation, we disregard
the minimum and maximum single-exposure values at each coadd pixel location. Precomputing
the minimum and maximum values at each coadd pixel location requires $\sim$0.01 seconds per
exposure, which is negligible relative to the overall runtime of coaddition, and typically leads to 
$\sim$10 additional seconds of CPU time per tile at low ecliptic latitude.

\begin{figure*}
	\includegraphics[width=7.0in]{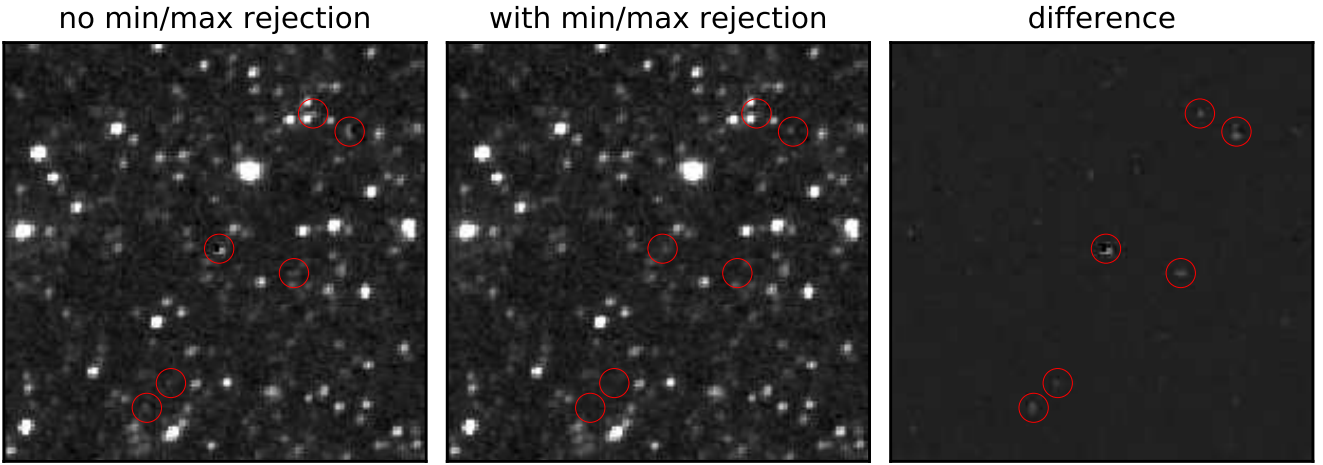}
    \caption{Benefits of min/max rejection for a W1 coadd affected by the South Atlantic Anomaly. 
     Left: coadd image created without min/max rejection step during first-round coaddition. Middle: coadd image created with 
     newly implemented min/max rejection step during first-round coaddition. Right: difference between left and center panels, highlighting spurious features associated with cosmic rays which have now 
     been removed. Red circles in each panel mark the locations of the most pronounced such artifacts.
     The region shown is $5.9'$$\times$$5.9'$ in size, and centered at 
     ($\alpha$, $\delta$) = (48.690$^{\circ}$, $-$23.923$^{\circ}$).}
    \label{fig:min_max_img_u}
\end{figure*}

The right panel of Figure \ref{fig:min_max_r1} shows an example of dramatically reduced cosmic ray 
contamination in a first round coadd mean image achieved by incorporating our min/max
rejection step. Obtaining a cleaner first-round estimate of the per-pixel mean and standard deviation
improves our ability to reject outliers during the second round of coaddition. This is illustrated
by the example region shown in Figure \ref{fig:min_max_img_u}, where 
min/max rejection has eliminated the imprint of cosmic ray strikes from our final coadd image.

Note that our min/max rejection step does not necessarily result in a reduction of integer coverage
for our final outputs at every coadd pixel, since it is only applied during the first round coaddition process. In future iterations of our full-depth coadds, as the typical integer coverage continues to increase, we 
may investigate more aggressive min/max rejection in which the highest/lowest $N$ single-exposure values at each coadd pixel location are discarded, with $N > 1$.

\subsection{Bitmasks}

One feature which has been absent from previous unWISE data releases is a pixel-level
bitmask corresponding to each coadd image. Such a bitmask data product could be very helpful for selection of spectroscopic samples (e.g. for DESI), in order to improve efficiency by flagging sources
associated with artifacts. It could also be quite useful in aiding rare object searches, by indicating
when unusual objects might in fact be of instrumental rather than astrophysical origin. For instance, a close
(W1-only) variant of the bitmask procedure presented here was used to great effect in the
Planet Nine search of \cite{p9w1}.

In this work, we address these needs by creating a pixel-level bright star mask output
for each coadd tile, with suffix \verb|-msk.fits.gz|. We construct one such bitmask image per 
tile, containing information about bright stars in both W1 and W2, rather than generating two 
separate bitmasks with one corresponding to each band.

The simplest bright star mask one might imagine would flag a circular region about every
sufficiently bright star, with the radius scaled based on each star's total flux according to
some empirically calibrated prescription. Indeed, the eBOSS collaboration currently employs masks 
of bright WISE stars created in this way \citep{eboss_lrg, eboss_qso}. Our bitmasks are designed to go beyond this simplistic approach, taking into account both the WISE PSF and the WISE scan direction to create highly detailed masks which include features such as diffraction spikes and optical ghosts.

As the basis for our bitmasks, we must select samples of very bright WISE sources in each of W1 and W2. We do so by making use of the positions and fluxes in the AllWISE catalog. We create a full-sky 
list of W1-bright sources by selecting objects from the AllWISE catalog with \verb|w1mpro| $<$ 9.5. 
For W2, we adopt a brightness threshold of \verb|w2mpro| $<$ 8.3. Our masks are primarily designed 
with extragalactic/cosmology applications in mind, and therefore are optimized for high Galactic latitude sky regions. In locations far enough off the Galactic plane to be part of DESI's footprint ($|b_{gal}| > 18^{\circ}$), our bright star sample contains on average $\sim$29 ($\sim$11) sources per square degree in W1 (W2).

For each \verb|coadd_id| astrometric footprint, we begin constructing its bitmask by 
identifying all bright stars which may have profiles at least partially falling inside of the
tile boundaries. We properly account for all bright stars which overlap the tile 
footprint of interest with any portion of their PSF wings/diffraction spikes,  even those for which the centroid falls outside of the tile boundaries. For each bright star, we then render a model of its
appearance given its AllWISE centroid and flux. To render these models, we make use of the
W1 and W2 PSFs of \cite{meisner14}. These PSFs, displayed in Figure 4 of \cite{lang14}, are $14.9'$ on a side, and therefore extend quite far into the wings, capturing details such as diffraction spikes and the W2 optical ghost.

One subtlety involved in rendering our bright star profiles is that the \cite{meisner14} PSFs provide models of the
single-exposure PSF, not the effective PSF obtained after resampling onto the unWISE tile
astrometric footprints and coadding. For instance, in L1b images, the diffraction spikes
always emanate from bright sources at (45$^{\circ}$, 135$^{\circ}$, 225$^{\circ}$, 315$^{\circ}$) degrees from the +$x_{L1b}$ direction. However, this is not the case for unWISE tiles, which are oriented
along the equatorial cardinal directions, whereas the L1b exposures are very nearly oriented exactly along the ecliptic cardinal directions.

A further subtlety is that the \cite{meisner14} PSF models require specification of detector 
($x_{L1b}$, $y_{L1b}$) coordinates, since they incorporate PSF variation across the 
single-exposure field of view (FOV). Therefore, before rendering any bright star models during bitmask construction, we precompute mean versions of the W1 and W2 \cite{meisner14} PSFs, averaging over ($x_{L1b}$, $y_{L1b}$) detector location.

\begin{figure}
	\includegraphics[width=3.3in]{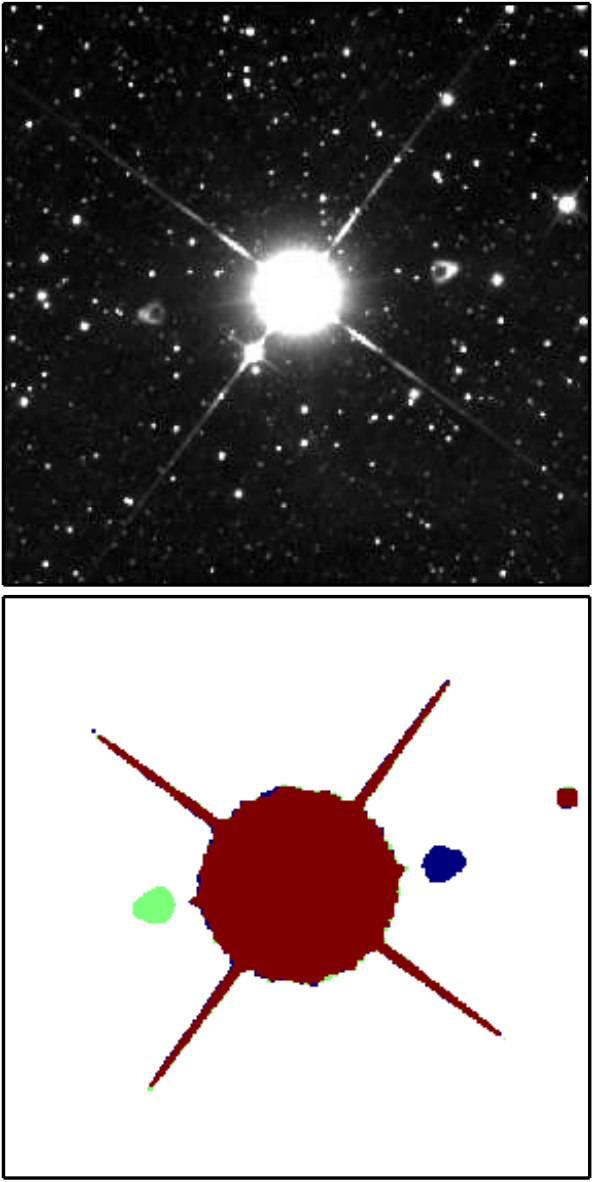}
    \caption{Illustration of our bitmask in the vicinity of an extremely bright star. Top: Grayscale rendering of our W2 full-depth coadd. Bottom: Colorscale rendering
    of our corresponding W2 mask bits. White indicates a mask value of 0. The masked regions corresponding to the two appearances of the doughnut-shaped ghost have different colors because they originate from opposite WISE scan
    directions. The region shown is 21.1$'$ $\times$ 21.1$'$ in size, centered at
    ($\alpha$, $\delta$) = (247.360$^{\circ}$, $-$19.347$^{\circ}$).}
    \label{fig:bitmask}
\end{figure}

A final subtlety is that, except for a small area near $|\beta| = 90^{\circ}$, there will
be two discrete WISE scan directions contributing to each coadd, corresponding to the ecliptic
north and south directions. The PSF is not perfectly symmetric with respect to swapping the
scan direction. The most notable asymmetry in the bands of interest is the W2 ghost, which
is offset by $\sim$5$'$ from its parent bright source centroid, at a position angle that is fixed relative to the scan direction. In practice this means that, in our coadds, the ghost will appear on opposite sides of its parent bright star in
exposures with opposite scan directions. This explains why the doughnut-shaped ghost in Figure \ref{fig:bitmask} appears twice.

In order to create a coadd-level model rendering of a bright star, we compute the angle between
the scan direction and celestial north, rotate the FOV-averaged PSF by this angle, and
scale the rotated profile so that its total flux matches that quoted by the AllWISE catalog. For each
bright star, we create two such renderings, one for each scan direction. Pixels in the 
bright star models above a threshold of 13.2 (38.3) Vega nanomaggies per square arcsecond in W1 (W2) are flagged according to the mask bits listed in Table
\ref{tab:bitmask}. At very high latitude, $|b_{gal}| > 80^{\circ}$, $\sim$0.4\% of the total sky area
is masked on average. This fraction ramps up toward lower $|b_{gal}|$, reaching 1.6\% at
$|b_{gal}|$ = 18$^{\circ}$, the closest DESI will observe to the Galactic plane.

\begin{table}
	\centering
	\caption{Description of bright star mask bits.}
	\label{tab:bitmask}
	\begin{tabular}{llc} 
		\hline
		Bit & Description &  Scan Direction\\
                 \hline
	        0 & W1 bright star & south \\
		1 & W1 bright star & north \\
		2 & W2 bright star & south \\
		3 & W2 bright star & north \\
		\hline
	\end{tabular}
\end{table}

Our pixel-level bitmasks have now been propagated into the DESI imaging
Legacy Survey DR4 catalogs. An example bitmask image near an extremely bright star is shown in Figure \ref{fig:bitmask}. Much room remains to extend our present bitmasks in future unWISE coadd releases, making them even more intricately detailed and elaborate. The primary avenues for doing so are:

\begin{itemize}
\item Near the Galactic plane, our current bitmasks flag an excessively large fraction of pixels, 
rendering the masks unhelpful in these regions. In the future, we could counteract this issue by making our \verb|w?mpro| thresholds dependent on e.g. Galactic latitude, so that a larger flux would be required to trigger masking of a star at low $b_{gal}$.
\item As seen in Figure \ref{fig:bitmask}, the 
diffraction spikes of extremely bright stars can sometimes extend
beyond the $14.9'$ boundary of the \cite{meisner14} PSF models. In the future, we could calibrate
look-up tables of diffraction spike length versus magnitude in W1 and W2, and use
these to create geometric masks composed of lines emanating 
from the centroids of extremely bright stars at angles of (45$^{\circ}$, 135$^{\circ}$, 225$^{\circ}$, 
315$^{\circ}$) relative to ecliptic north. Such a procedure was employed in W3 by \cite{meisner14}, as
described in their $\S$5.2.4.
\item ``Latents'' represent another class of defects associated with stars bright enough to saturate in their cores. Latents are persistence artifacts which appear as diffuse positive blobs at the 
detector positions of saturated pixels, but in the frame immediately after imaging of the 
parent bright star. The locations of all latents can be predicted exactly given L1b metadata and a catalog of bright stars, so it would be possible to reliably flag these artifacts by adding new mask bits.
\item Our assumption that there are exactly two discrete scan directions corresponding to
ecliptic north and south is violated at the ecliptic poles, where continuous coverage yields a continuum of scan directions. As a result, very near $|\beta|$=90$^{\circ}$, bright star diffraction spikes become
spread out into discs in the coadds and are correspondingly attenuated. One approach for taking these effects into account has been described in $\S$5.2.4 of \cite{meisner14}.
\end{itemize}

\section{Results}
\label{sec:results}

\subsection{Images}

\begin{figure}
	\includegraphics[width=3.3in]{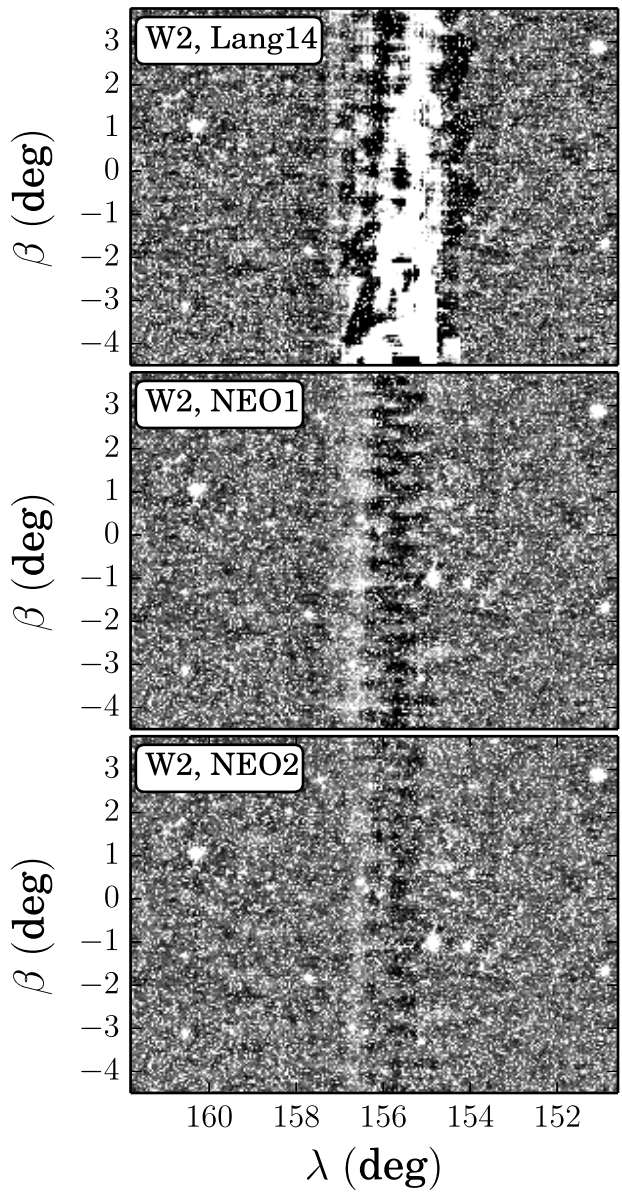}
    \caption{Continuing reduction of W2 scattered moonlight contamination thanks to
    the added redundancy of second-year NEOWISER exposures. Top: Low-resolution rendering
    of an $11.2^{\circ} \times 8.3^{\circ}$ region near the ecliptic plane, based on the \citet{lang14}
    unWISE coadds. In W1/W2, these coadds had no special handling of scattered moonlight, 
    resulting in a series of significantly corrupted vertical streaks, one of which is shown here. 
    Middle: same region in the \citet{meisner16} W2 coadds. The major improvement is due to 
    the adoption of additional outlier rejection for Moon-contaminated frames, plus added redundancy 
    from incorporating first-year NEOWISER images. Bottom: Further reduction of the 
    Moon contamination is apparent in the W2 coadds of this work, thanks to our inclusion of 
    second-year NEOWISER images, which essentially averages down any remaining Moon-related 
    artifacts. The grayscale stretch is identical in all cases, ranging linearly from $-0.4$ (black) to 
    1.5 (white) Vega nanomaggies per square arcsecond.}
    \label{fig:moon}
\end{figure}

\begin{figure*}
	\includegraphics[width=7.3in]{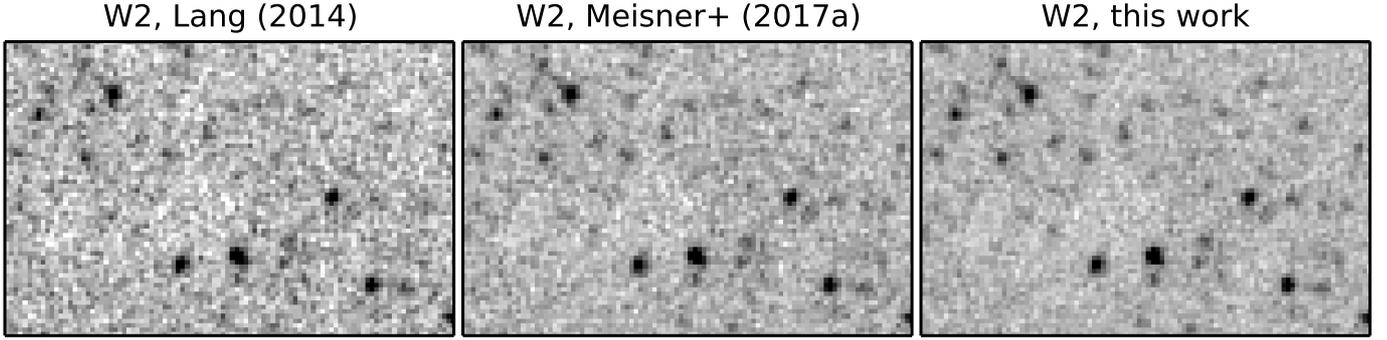}
    \caption{Typical reduction of image noise with
    increasing depth of coverage at low ecliptic latitude. Left: \citet{lang14} unWISE coadd, based on pre-hibernation observations. Center: \citet{meisner16} coadd, including both pre-hibernation and first-year NEOWISER exposures. Right: This work, additionally incorporating second-year NEOWISER frames. The region shown is 4.6$'$ $\times$ 3.0$'$ in size, centered at ($\alpha$, $\delta$) = (150.171$^{\circ}$, 1.021$^{\circ}$).}
    \label{fig:depth}
\end{figure*}

Relative to the coadds of \cite{meisner16}, we have in general increased the depth of coverage by $\sim$50\% while further mitigating artifacts.
The mean integer coverage over the entire sky is 108 (107) frames in W1 (W2), and every
pixel has integer coverage of at least 33 (30) in W1 (W2). Figure \ref{fig:moon} illustrates that folding in a third full year of observations has reduced the impact of scattered moonlight, the dominant systematic problem with the W1/W2 unWISE coadds on large angular scales. Because the Moon affects
different ranges of ecliptic longitude during different sky passes, any remaining traces of scattered moonlight contamination become further suppressed as additional NEOWISER coverage is incorporated. Figure \ref{fig:depth} shows a typical example of the reduction in statistical noise achieved for coadds at low ecliptic latitude.

\subsection{Catalogs}

DR4 of the DESI imaging Legacy Survey, hereafter referred to as simply `DR4', contains the only currently available photometry based on the  full-depth W1/W2 coadds presented in this work. This photometry is forced rather than WISE-selected, adopting source locations and morphologies derived from deep optical data. DR4 covers a
$\sim$4,500 square degree extragalactic footprint limited to $\delta \gtrsim +30^{\circ}$, as
shown in Figure 3.19 of \cite{desi_part1}.

In validating our coadds, we first seek catalog-level confirmation of the decrease in statistical noise suggested by the images in Figure \ref{fig:depth}. We examine objects with W1/W2 forced photometry available from both \cite{lang14b}, based on the original unWISE coadds, and DR4 of the Legacy Survey. We select a comparison sample drawn from
the region $170^{\circ} < \alpha < 230^{\circ}$, $42.5^{\circ} < \delta < 48^{\circ}$, using a 1$''$ match radius, and restricting to objects with DR4 forced photometry signal-to-noise in the range 10 $\pm$ 1.
For these 127,000 (144,000) sources in W1 (W2), we find that forced photometry
flux uncertainties from Legacy Survey DR4 are smaller than those of \cite{lang14b} by median factors of 1.68$\times$ (1.53$\times$). These values indicate that forced photometry based on the 
present coadds is 0.56 (0.46) mags deeper in W1 (W2) than forced photometry based solely on pre-hibernation W1/W2 imaging. For both the W1 and W2 samples, the median increase in integer
coverage is a factor of 2.94$\times$, leading us to expect flux uncertainties reduced by 1.71$\times$ under the assumption that all WISE observations have maintained the same sensitivity regardless of mission phase.

\begin{figure*}
	\includegraphics[width=6.0in]{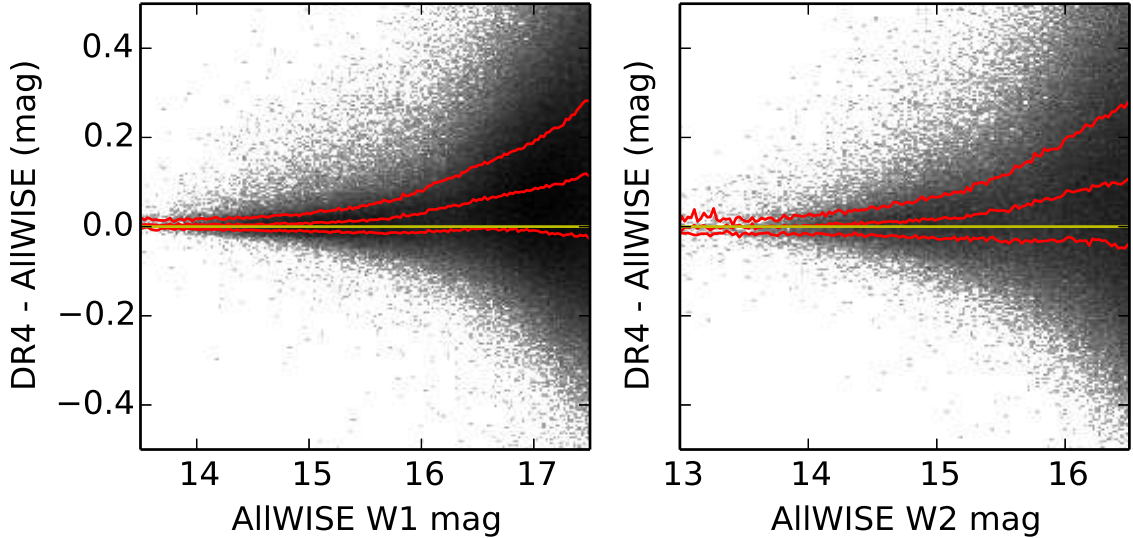}
    \caption{Comparison of forced photometry based on our new W1/W2 coadds against flux measurements from the AllWISE catalog.  The forced photometry is obtained from DR4 of the DESI imaging Legacy Survey. Left: W1 residual versus AllWISE W1 magnitude. Right: W2 residual versus AllWISE W2 magnitude.}
    \label{fig:allwise}
\end{figure*}

We additionally check that forced photometry fluxes derived from our new coadds are consistent with the AllWISE catalog. We select a comparison sample drawn from
the region $170^{\circ} < \alpha < 230^{\circ}$, $42.5^{\circ} < \delta < 48^{\circ}$, using a 1$''$ match radius. We further require a
one-to-one DR4-AllWISE match, DR4 morphological type \verb|PSF| and \verb|w?cc_map| = 0 in the band of interest. This yields 626,000 (612,000) sources in
W1 (W2). Figure \ref{fig:allwise} summarizes our DR4-AllWISE comparison for this sample. There is good overall agreement across a wide range of fluxes, and the (DR4 $-$ AllWISE) offset asymptotes to 4.1 ($-$1.6) mmag in W1 (W2) toward the bright end. The
faint end upturn has previously been noted in \cite{lang14b} and \cite{meisner16}, and is due to Malmquist bias, as the AllWISE sources are WISE-selected while the DR4 objects are not.

\subsection{Data Access}
The full-depth W1/W2 coadds described in this work are publicly available via the unWISE web interface at \url{http://unwise.me}.

\section{Conclusion}
\label{sec:conclusion}
We have reprocessed all publicly available W1/W2 observations ever acquired to 
create the deepest full-sky maps at 3$-$5$\mu$m.  New processing steps have been
introduced, focused on enabling clean target selection and rare object searches. We have also validated our new coadds using forced photometry from DR4 of the DESI imaging Legacy Survey. It will
be important to continue updating our full-depth W1/W2 stacks with each future release of additional NEOWISER exposures, as doing so represents a crucial step toward realizing the full potential of the entire WISE imaging data set.

\section*{Acknowledgements}
This work has been supported by grant NNH17AE75I from the NASA Astrophysics Data Analysis Program.

This research makes use of data products from the Wide-field Infrared Survey Explorer, which is a joint project of the University of California, Los Angeles, and the Jet Propulsion Laboratory/California Institute of Technology, funded by the National Aeronautics and Space Administration. This research also makes use of data products from NEOWISE, which is a project of the Jet Propulsion Laboratory/California Institute of Technology, funded by the Planetary Science Division of the National Aeronautics and Space Administration. This research has made use of the NASA/ IPAC Infrared Science Archive, which is operated by the Jet Propulsion Laboratory, California Institute of Technology, under contract with the National Aeronautics and Space Administration.

The National Energy Research Scientific Computing Center, which is supported by the Office of Science of the U.S. Department of Energy under Contract No. DE-AC02-05CH11231, provided staff, computational resources, and data storage for this project.

\bibliographystyle{mnras}
\bibliography{fulldepth_neo2} 

\begin{thebibliography}{}
\makeatletter
\relax
\def\mn@urlcharsother{\let\do\@makeother \do\$\do\&\do\#\do\^\do\_\do\%\do\~}
\def\mn@doi{\begingroup\mn@urlcharsother \@ifnextchar [ {\mn@doi@}
  {\mn@doi@[]}}
\def\mn@doi@[#1]#2{\def\@tempa{#1}\ifx\@tempa\@empty \href
  {http://dx.doi.org/#2} {doi:#2}\else \href {http://dx.doi.org/#2} {#1}\fi
  \endgroup}
\def\mn@eprint#1#2{\mn@eprint@#1:#2::\@nil}
\def\mn@eprint@arXiv#1{\href {http://arxiv.org/abs/#1} {{\tt arXiv:#1}}}
\def\mn@eprint@dblp#1{\href {http://dblp.uni-trier.de/rec/bibtex/#1.xml}
  {dblp:#1}}
\def\mn@eprint@#1:#2:#3:#4\@nil{\def\@tempa {#1}\def\@tempb {#2}\def\@tempc
  {#3}\ifx \@tempc \@empty \let \@tempc \@tempb \let \@tempb \@tempa \fi \ifx
  \@tempb \@empty \def\@tempb {arXiv}\fi \@ifundefined
  {mn@eprint@\@tempb}{\@tempb:\@tempc}{\expandafter \expandafter \csname
  mn@eprint@\@tempb\endcsname \expandafter{\@tempc}}}

\bibitem[\protect\citeauthoryear{{Boggess} et~al.,}{{Boggess}
  et~al.}{1992}]{dirbe}
{Boggess} N.~W.,  et~al., 1992, \mn@doi [\apj] {10.1086/171797}, \href
  {http://adsabs.harvard.edu/abs/1992ApJ...397..420B} {397, 420}

\bibitem[\protect\citeauthoryear{{Connors}, {Wiegert}  \& {Veillet}}{{Connors}
  et~al.}{2011}]{2010TK7}
{Connors} M.,  {Wiegert} P.,   {Veillet} C.,  2011, \mn@doi [\nat]
  {10.1038/nature10233}, \href
  {http://adsabs.harvard.edu/abs/2011Natur.475..481C} {475, 481}

\bibitem[\protect\citeauthoryear{{Cutri} et~al.,}{{Cutri}
  et~al.}{2012}]{cutri12}
{Cutri} R.~M.,  et~al., 2012, Technical report, {Explanatory Supplement to the
  WISE All-Sky Data Release Products}

\bibitem[\protect\citeauthoryear{{Cutri} et~al.,}{{Cutri}
  et~al.}{2013}]{cutri13}
{Cutri} R.~M.,  et~al., 2013, Technical report, {Explanatory Supplement to the
  AllWISE Data Release Products}

\bibitem[\protect\citeauthoryear{{Cutri} et~al.,}{{Cutri}
  et~al.}{2015}]{cutri15}
{Cutri} R.~M.,  et~al., 2015, Technical report, {Explanatory Supplement to the
  NEOWISE Data Release Products}

\bibitem[\protect\citeauthoryear{{DESI Collaboration} et~al.,}{{DESI
  Collaboration} et~al.}{2016a}]{desi_part1}
{DESI Collaboration} et~al., 2016a, preprint, \href
  {http://adsabs.harvard.edu/abs/2016arXiv161100036D} {} (\mn@eprint {arXiv}
  {1611.00036})

\bibitem[\protect\citeauthoryear{{DESI Collaboration} et~al.,}{{DESI
  Collaboration} et~al.}{2016b}]{desi_part2}
{DESI Collaboration} et~al., 2016b, preprint, \href
  {http://adsabs.harvard.edu/abs/2016arXiv161100037D} {} (\mn@eprint {arXiv}
  {1611.00037})

\bibitem[\protect\citeauthoryear{{Faherty} et~al.,}{{Faherty}
  et~al.}{2015}]{faherty}
{Faherty} J.~K.,  et~al., 2015, preprint, \href
  {http://adsabs.harvard.edu/abs/2015arXiv150501923F} {} (\mn@eprint {arXiv}
  {1505.01923})

\bibitem[\protect\citeauthoryear{{Lang}}{{Lang}}{2014}]{lang14}
{Lang} D.,  2014, \mn@doi [\aj] {10.1088/0004-6256/147/5/108}, \href
  {http://adsabs.harvard.edu/abs/2014AJ....147..108L} {147, 108}

\bibitem[\protect\citeauthoryear{{Lang}, {Hogg}  \& {Schlegel}}{{Lang}
  et~al.}{2016}]{lang14b}
{Lang} D.,  {Hogg} D.~W.,   {Schlegel} D.~J.,  2016, \mn@doi [\aj]
  {10.3847/0004-6256/151/2/36}, \href
  {http://adsabs.harvard.edu/abs/2016AJ....151...36L} {151, 36}

\bibitem[\protect\citeauthoryear{{Levi} et~al.,}{{Levi} et~al.}{2013}]{desi}
{Levi} M.,  et~al., 2013, preprint, \href
  {http://adsabs.harvard.edu/abs/2013arXiv1308.0847L} {} (\mn@eprint {arXiv}
  {1308.0847})

\bibitem[\protect\citeauthoryear{{Mainzer} et~al.,}{{Mainzer}
  et~al.}{2011}]{neowise}
{Mainzer} A.,  et~al., 2011, \mn@doi [\apj] {10.1088/0004-637X/731/1/53}, \href
  {http://adsabs.harvard.edu/abs/2011ApJ...731...53M} {731, 53}

\bibitem[\protect\citeauthoryear{{Mainzer} et~al.,}{{Mainzer}
  et~al.}{2014}]{neowiser}
{Mainzer} A.,  et~al., 2014, \mn@doi [\apj] {10.1088/0004-637X/792/1/30}, \href
  {http://adsabs.harvard.edu/abs/2014ApJ...792...30M} {792, 30}

\bibitem[\protect\citeauthoryear{{Meisner} \& {Finkbeiner}}{{Meisner} \&
  {Finkbeiner}}{2014}]{meisner14}
{Meisner} A.~M.,  {Finkbeiner} D.~P.,  2014, \mn@doi [\apj]
  {10.1088/0004-637X/781/1/5}, \href
  {http://adsabs.harvard.edu/abs/2014ApJ...781....5M} {781, 5}

\bibitem[\protect\citeauthoryear{{Meisner}, {Lang}  \& {Schlegel}}{{Meisner}
  et~al.}{2017a}]{meisner16}
{Meisner} A.~M.,  {Lang} D.,   {Schlegel} D.~J.,  2017a, \mn@doi [\aj]
  {10.3847/1538-3881/153/1/38}, \href
  {http://adsabs.harvard.edu/abs/2017AJ....153...38M} {153, 38}

\bibitem[\protect\citeauthoryear{{Meisner}, {Bromley}, {Nugent}, {Schlegel},
  {Kenyon}, {Schlafly}  \& {Dawson}}{{Meisner} et~al.}{2017b}]{p9w1}
{Meisner} A.~M.,  {Bromley} B.~C.,  {Nugent} P.~E.,  {Schlegel} D.~J.,
  {Kenyon} S.~J.,  {Schlafly} E.~F.,   {Dawson} K.~S.,  2017b, \mn@doi [\aj]
  {10.3847/1538-3881/153/2/65}, \href
  {http://adsabs.harvard.edu/abs/2017AJ....153...65M} {153, 65}

\bibitem[\protect\citeauthoryear{{Myers} et~al.,}{{Myers}
  et~al.}{2015}]{eboss_qso}
{Myers} A.~D.,  et~al., 2015, \mn@doi [\apjs] {10.1088/0067-0049/221/2/27},
  \href {http://adsabs.harvard.edu/abs/2015ApJS..221...27M} {221, 27}

\bibitem[\protect\citeauthoryear{{Prakash}, {Licquia}, {Newman}  \&
  {Rao}}{{Prakash} et~al.}{2015}]{prakash15}
{Prakash} A.,  {Licquia} T.~C.,  {Newman} J.~A.,   {Rao} S.~M.,  2015, \mn@doi
  [\apj] {10.1088/0004-637X/803/2/105}, \href
  {http://adsabs.harvard.edu/abs/2015ApJ...803..105P} {803, 105}

\bibitem[\protect\citeauthoryear{{Prakash} et~al.,}{{Prakash}
  et~al.}{2016}]{eboss_lrg}
{Prakash} A.,  et~al., 2016, \mn@doi [\apjs] {10.3847/0067-0049/224/2/34},
  \href {http://adsabs.harvard.edu/abs/2016ApJS..224...34P} {224, 34}

\bibitem[\protect\citeauthoryear{{Schlegel} et~al.,}{{Schlegel}
  et~al.}{2015}]{decals}
{Schlegel} D.~J.,  et~al., 2015, in American Astronomical Society Meeting
  Abstracts. p. 336.07

\bibitem[\protect\citeauthoryear{{Tsai} et~al.,}{{Tsai} et~al.}{2015}]{tsai15}
{Tsai} C.-W.,  et~al., 2015, \mn@doi [\apj] {10.1088/0004-637X/805/2/90}, \href
  {http://adsabs.harvard.edu/abs/2015ApJ...805...90T} {805, 90}

\bibitem[\protect\citeauthoryear{{Wheelock} et~al.,}{{Wheelock}
  et~al.}{1994}]{iras}
{Wheelock} S.~L.,  et~al., 1994, NASA STI/Recon Technical Report N, \href
  {http://adsabs.harvard.edu/abs/1994STIN...9522539W} {95}

\bibitem[\protect\citeauthoryear{{Wright} et~al.,}{{Wright}
  et~al.}{2010}]{wright10}
{Wright} E.~L.,  et~al., 2010, \mn@doi [\aj] {10.1088/0004-6256/140/6/1868},
  \href {http://adsabs.harvard.edu/abs/2010AJ....140.1868W} {140, 1868}

\makeatother
\end{thebibliography}

\label{lastpage}
\end{document}